\documentclass[letter]{aa}

\usepackage{txfonts}
\usepackage{amsmath,amssymb}

\usepackage{hyperref}
\hypersetup{
    colorlinks,
	unicode,
    linkcolor={red!80!black},
    citecolor={blue!80!black},
    urlcolor={blue!80!black}
}

\usepackage{natbib}

\newcommand{\kima}{\texttt{kima}\xspace}

\newcommand{\fR}{\ensuremath{f_R}\xspace}

\newcommand{\Ponly}{\ensuremath{\mathcal{P}}\xspace}
\newcommand{\ms}{\ensuremath{\mathrm{m} \, \mathrm{s}^{-1}}\xspace}
\newcommand{\given}[1][]{\:#1\vert\:}

\begin{document}

\title{Inferring planet occurrence rates from radial velocities}

\author{
  J.~P.~Faria\thanks{Corresponding author: joao.faria@unige.ch} \and
  J.-B.~Delisle \and
  D.~Ségransan
}

\institute{
  Observatoire Astronomique de l'Université de Genève, Chemin Pegasi 51b, 1290 Versoix, Switzerland \label{ inst1 } 
}

\date{\today}

\abstract%
{ %
	We introduce a new method to infer the posterior distribution for planet
	occurrence rates from radial-velocity (RV) observations. The approach combines
	posterior samples from the analysis of individual RV datasets of several
	stars, using importance sampling to reweight them appropriately. This
	eliminates the need for injection-recovery tests to compute detection limits
	and avoids the explicit definition of a detection threshold.
	We validate the method on simulated RV datasets and show that it yields
	unbiased estimates of the occurrence rate in different regions, with
	increasing precision as more stars are included in the analysis.
}



\maketitle

\section{Introduction}

The observed population of exoplanets provides a direct probe into the complex
processes governing the formation and evolution of planetary systems, although
it is inevitably affected by the biases of the different detection techniques.
For instance, the radial-velocity (RV) method is most sensitive to massive
planets with intermediate orbital periods in comparison to the transit or
direct imaging methods. \citep[e.g.][]{wright2013}.

An important goal of many RV surveys is to characterise the planet occurrence
rates across different stellar populations \citep[e.g.,][and references
therein]{cumming1999, howard2010, bonfils2013, rosenthal2021}. Achieving this
requires careful sample selection, accurate stellar characterization, detailed
analysis of the individual RV time series, and an assessment of survey
completeness to correct for observational biases.

In this Letter, we present a simple method to combine results from independent
RV analyses into a probabilistic estimate of the occurrence rate of planets
within a given region of parameter space. The approach bypasses the explicit
definition of detection thresholds and the calculation of detection limits or
survey completeness, relying only on posterior samples from existing analyses.
This allows for efficiently estimating occurrence rates in different regions
and stellar samples.

Section \ref{sec:methods} reviews the Bayesian approach to RV analysis and
outlines the proposed method. In Section \ref{sec:simulations}, we present an
application of the method to synthetic RV datasets, and in Section
\ref{sec:discussion} we discuss the results.

\section{Methods}\label{sec:methods}

\subsection{Bayesian RV analysis}

The use of Bayesian methods in the analysis of RVs is now an almost standard
practice. The seminal works of \citet{Ford2005a} and \citet{Gregory2005}
provided the general framework for the use of Markov chain Monte Carlo
techniques to sample the posterior distribution of the model parameters. More
recently, Nested Sampling \citep{Skilling2004a,Skilling2006} and several
variants \citep[e.g.][]{higson2019,feroz2009,buchner2023} have been widely
used for more efficient sampling and to allow for the comparison of competing
models.

The Bayesian approach boils down to a transformation of a prior distribution
$\pi(\theta \given \mathcal{M})$ for the parameters $\theta$ of a model
$\mathcal{M}$ into a posterior distribution $p(\theta \given \mathcal{D},
\mathcal{M})$, after having observed data $\mathcal{D}$. Bayes' theorem relates
these distributions as
\begin{equation}
	p \left( \mathcal{D} \given \mathcal{M} \right)
	\cdot
	p \left( \theta \given \mathcal{D}, \mathcal{M} \right)
	=
	p \left( \mathcal{D} \given \theta, \mathcal{M} \right)
	\cdot
	\pi \left( \theta \given \mathcal{M} \right)
	\, ,
\end{equation}
where we have introduced the likelihood $p\left(\mathcal{D} \given \theta,
	\mathcal{M}\right)$ and the evidence $p\left(\mathcal{D} \given
	\mathcal{M}\right)$, often denoted simply by $\mathcal{Z}$. Writing Bayes
theorem in this form emphasises the inputs (prior and likelihood) and outputs
(evidence and posterior) of any Bayesian analysis \citep[see][for a
	discussion]{Skilling2004a}.

In the RV analysis context, the number of orbiting planets in a given system,
$N_p$, is an interesting quantity.
Often, the value of $N_p$ is fixed in sequence and the evidence is used to
compare the models and estimate the number of planets significantly detected
in the dataset \citep[e.g.][]{feroz2011a}. This requires the definition of a
threshold for what is considered a significant detection, or a significant
difference in evidence. Although much literature has been dedicated to this
topic, the interpretation scale introduced by \citet{jeffreys1939} and refined
by \citet{kass1995} is very often used in the exoplanet community.

Alternatively, $N_p$ itself can be included in the set of model parameters and
its posterior distribution estimated directly. Since $N_p$ is a discrete
parameter and it controls the dimensionality of the model, certain algorithms
(often called trans-dimensional) are more adequate for this task
\citep[e.g.][]{green1995,brewer2014}. Such methods were first applied to RV
analysis by \citet{brewer2015} and have since been used to characterise
several exoplanet systems \citep[e.g.][]{faria2016, lillo-box2020, faria2022,
	baycroft2023, john2023}. The posterior for $N_p$ allows for efficient model
comparison, in particular when several planets are detected.

\subsection{Compatibility limits}

A typical output from the RV analysis of a particular system, whether it led to
planet detections or not, is an upper limit on the mass of planets that can be
ruled out given the observations. These detection limits are almost always
estimated by injecting mock signals into a residual dataset and assessing their
recovery, usually with a periodogram analysis
\citep[e.g.][]{cumming1999,mayor2011,bonfils2013a}.

This injection-recovery procedure is not well justified statistically, and can
be costly computationally. Moreover, it requires the definition of a threshold
for detection and the subtraction of a single solution from the original
dataset, therefore ignoring any uncertainty and correlations present in the
posterior.

But in fact, the posterior distribution may already contain all the available
information about which orbital parameters are compatible with the data. For
example, if no signals are detected, all posterior samples from models with $N_p
\geq 1$ are compatible with the data and can be used to estimate so-called
compatibility limits \citep[][see also \citealt{tuomi2014a}]{standing2022,
figueira2025}, akin to the detection limits. If instead one signal is detected,
posterior samples with $N_p \geq 2$ can be used.

This approach has several advantages. First, compatibility limits are available
from a single run of the sampling algorithm, provided the prior for $N_p$ allows
for sufficiently large values. Second, orbital parameters such as eccentricity
can be effectively marginalised over, subject to the prior distribution.
Finally, the compatibility limits can be calculated without subtracting any
single solution from the original data.

While useful in providing an estimated sensitivity of an RV dataset to planets
of different periods and masses, which could be combined into a survey
sensitivity map, the compatibility limits are not required for determining the
occurrence rate.

\subsection{Planet occurrence rates}
\label{subsec:occurrence}

The analysis of RV datasets of multiple stars provides information about the
occurrence rate of planets. We define occurrence rate as the fraction of stars
with at least one planet in a region $R$ of the parameter space%
\footnote{But see Appendix \ref{sec:appD} for an alternative definition.}. %
The region can be defined in different ways, taking different planet properties
into account. We will only consider lower and upper limits in orbital period and
in planetary minimum mass. The occurrence rate will generally depend on the
choice of the region.

In probabilistic terms, we define the event ``there is at least one planet with
parameters in a given region''. This event, here denoted as \Ponly, takes values
1 or 0 (true or false),
\begin{equation}
	\Ponly \equiv
	\begin{cases}
		1 \quad \text{ if } \, \theta \in R \\
		0 \quad \text{ otherwise}
	\end{cases}
\end{equation}
and, for each individual star, follows a Bernoulli distribution with probability
equal to the occurrence rate $\fR$, so that
\begin{equation}\label{eq:bernoulli}
	p \left( \fR; \Ponly \right)
	= \fR \cdot \Ponly + \left(1 - \fR\right) \cdot \left(1 - \Ponly\right)
	.
\end{equation}

For a total of $S$ stars (indexed by $j$), each with an RV dataset
$\mathcal{D}_j$, we have $M_j$ samples from the posterior $p(\theta_j \given
\mathcal{D}_j)$, obtained from the likelihood $p(\mathcal{D}_j \given \theta_j)$
and prior $\pi(\theta_j)$, where we omitted the dependence on the model
$\mathcal{M}$. If the observations of each star are independent, the total
likelihood for all the parameters of all stars is equal to the product of the
individual-star likelihoods:
\begin{equation}\label{eq:total_likelihood_0}
	p \left(
	\left\{ \mathcal{D}_j \right\}_S \given[\big] \left\{ \theta_j \right\}_S
	\right)
	= \prod_{j=1}^S p \left( \mathcal{D}_j \given \theta_j \right),
\end{equation}
where the notation $\{.\}_a$ means the set of $a$ elements. We comment on the
independence assumption in Section \ref{sec:discussion}.

We are interested in the posterior distribution (and thus in the likelihood) in
terms of the occurrence rate \fR, which will imply marginalising out the
individual-star parameters
\begin{equation}\label{eq:total_likelihood}
	p \left(
	\left\{ \mathcal{D}_j \right\}_S \given[\big] \fR
	\right)
	=
	\prod_{j=1}^S \int d\theta_j \cdot p \left( \mathcal{D}_j \given \theta_j \right)
	\cdot
	p \left( \theta_j \given \fR \right)
	.
\end{equation}

An issue arises because we do not have access to $p(\theta_j \given \fR)$ from
the individual analyses, since the occurrence rate was not taken into account
when computing the posteriors%
\footnote{In particular, the individual analyses need not consider any
	specific region $R$ where the occurrence rate is to be estimated.}%
\!. %
In order to change variables without having to redo the individual analyses (for
each different region), we can employ importance sampling to re-weight the
original prior \citep[e.g.][]{hogg2010a}:
\begin{equation}\label{eq:is}
	p \left( \theta_j \given \fR \right)
	\equiv
	\frac{ p(\fR; \Ponly_j) \cdot \pi(\theta_j) }{ \pi(\Ponly_j) }
	.
\end{equation}

Here, $\pi(\Ponly_j)$ is the distribution of $\Ponly_j$ under the original
prior.
Substituting Eq. \eqref{eq:is} into Eq. \eqref{eq:total_likelihood} results in
integrals over the posterior distribution $p (\mathcal{D}_j \given \theta_j)
\cdot \pi(\theta_j)$ multiplied by $p(\fR; \Ponly_j) / \pi(\Ponly_j)$. These
integrals can be approximated using averages of the posterior samples from each
star, so the likelihood becomes
\begin{equation}\label{eq:is_likelihood}
	p \left( \left\{ \mathcal{D}_j \right\}_S \given[\big] \fR \right)
	\approx
	\prod_{j=1}^S
	\frac{1}{M_j} \sum_{k=1}^{M_j} \frac{ p(\fR; \Ponly_{jk}) }{ \pi(\Ponly_{jk}) }
	,
\end{equation}

For the original prior, the equivalent to Eq. \eqref{eq:bernoulli} is
\begin{equation}\label{eq:bernoulli_f0}
	\pi (f_0; \Ponly_j) = f_0 \cdot \Ponly_j + (1 - f_0) \cdot (1 - \Ponly_j)
\end{equation}
with $f_0 \neq \fR$. The value of $f_0$ can be estimated (for each star) with a
simple procedure, outlined in Appendix \ref{sec:appA}. Then, we can write each
$j$-th average in Eq. \eqref{eq:is_likelihood} as
\begin{equation}
	\frac{\fR}{f_0} \cdot p(\Ponly_j \given \mathcal{D}_j)
	+
	\frac{1 - \fR}{1 - f_0} \cdot (1 - p(\Ponly_j \given \mathcal{D}_j))
	,
\end{equation}
where $p(\Ponly_j \given \mathcal{D}_j)$ is the posterior probability of
$\Ponly_j$, obtained simply as the fraction of posterior samples that fall
within the region $R$.
Combining the expression for the total likelihood with a prior for \fR, we
obtain the posterior distribution for the occurrence rate given all the
individual datasets:
\begin{equation}
	p \left(
	\fR \given[\big] \left\{ \mathcal{D}_j \right\}_S, \left\{ \theta_j \right\}_S
	\right)
	\propto
	p(\fR)
	\cdot
	p \left( \left\{ \mathcal{D}_j \right\}_S \given[\big] \fR \right)
	.
\end{equation}

The prior $p(\fR)$ can be an uninformative uniform distribution between 0 and 1.
The posterior can then be evaluated on a grid of values for \fR. Intuitively,
these expressions tell us that when the individual posterior probability within
a given region is high compared to what is expected from the prior, the
occurrence rate within that region is likely to be high.

\subsection{Implementation in \kima}

In this work, we use the \kima package \citep{kima}%
\footnote{Available at \href{https://github.com/kima-org/kima}{github.com/kima-org/kima}.} %
for the analysis of the RV datasets. By default, the code models the RVs with a
sum of Keplerian functions, sampling from the joint posterior distribution with
the Diffusive Nested Sampling algorithm \citep[DNS;][]{brewer2011}, which also
provides the evidence $\mathcal{Z}$. In addition, DNS supports sampling in the
trans-dimensional models mentioned before \citep{brewer2015}, in which the
number of planets is a free parameter. After an appropriate prior is defined,
the posterior distribution for $N_p$ is estimated together with the other model
parameters.

The resulting samples from the posterior distributions for several stars can be
directly used to compute the posterior for the occurrence rate, as described in
the previous section. Even if the calculation is relatively simple, we provide a
Python implementation of the method for convenience%
\footnote{See \href{https://github.com/kima-org/occurrence}{github.com/kima-org/occurrence}.}%
\!.

\section{Simulations}\label{sec:simulations}

To demonstrate the proposed approach, we generated a total of 50 synthetic RV
datasets (from 50 hypothetical stars), each containing between 40 and 50
measurements, drawn uniformly over a timespan of one year. We define three
regions of parameter space on which we want to estimate the occurrence rate:
\begin{itemize}
	\item[$R_1$:] $P \in (2, 25)$ d;    $m \in (3, 30)\,M_{\oplus}$   with assumed $\fR=0.3$
	\item[$R_2$:] $P \in (60, 100)$ d;  $m \in (50, 200)\,M_{\oplus}$ with assumed $\fR=0.7$
	\item[$R_3$:] $P \in (100, 400)$ d; $m \in (1, 10)\,M_{\oplus}$   with assumed $\fR=0.2$
\end{itemize}

\begin{figure}
	\centering
	\includegraphics[width=\hsize]{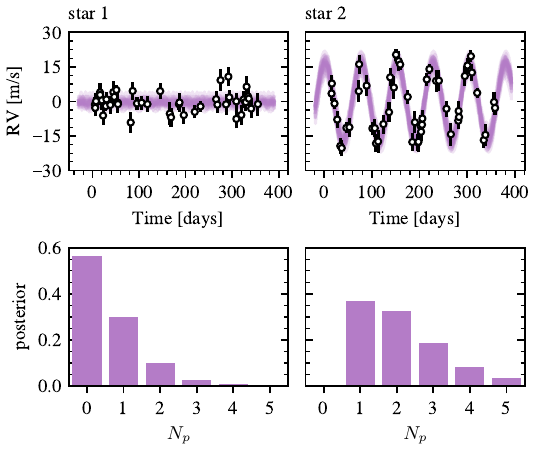}
	\caption{Example of two simulated datasets, containing no planet signal
		(left) and two planet signals (right). The top panels show the RV timeseries
		and random samples from the posterior predictive distribution. The bottom
		panels show the posterior distributions for $N_p$.}
	\label{fig:examples}
\end{figure}

\begin{figure}
	\centering
	\includegraphics[width=\hsize]{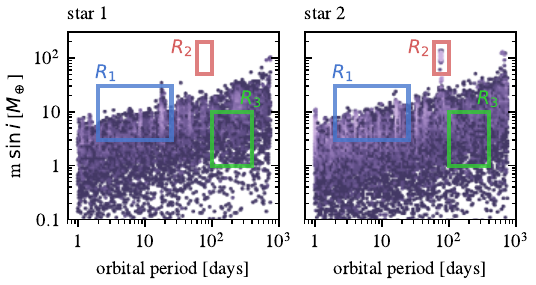}
	\caption{Joint posteriors for the minimum mass and orbital period, shown as
		     hexbin density plots, for the same example datasets as in
		     Fig.~\ref{fig:examples}. All samples with $N_p>0$ are shown. The
		     three regions where the occurrence rates are estimated are shown as
		     coloured boxes and labels.}
	\label{fig:examples_PK}
\end{figure}

For each star and for each of these three regions, we generate one mock planet
with probability equal to the corresponding \fR, and draw its orbital parameters
from the prior $\pi(\theta)$, constrained to the region. That is, each RV
timeseries can have between zero and three simulated Keplerian signals, with
parameters within the three regions.

All mock stars are assumed to have a mass of 1 $M_\odot$ and zero systemic
velocity. We draw RV uncertainties for each measurement from a uniform
distribution in variance, between $2^2$ and $5^2$ m$^2\,s^{-2}$ and add white
Gaussian noise to the RVs following these heteroscedastic variances. No
additional noise is added, but a jitter parameter is still used in the analysis.
Two examples of the generated datasets are shown in the top panels of
Fig.~\ref{fig:examples}.

We perform the analysis of each dataset with \kima, using default settings for
the DNS algorithm. The priors used in the analysis are listed in
Table~\ref{tab:priors} of Appendix \ref{sec:appB}. We note that the prior for
the orbital period extends to two years, even though the time spans of the
datasets are around one year. The DNS algorithm ran for 100\,000 iterations,
providing more than 10\,000 effective posterior samples for each dataset.

The bottom panels of Fig.~\ref{fig:examples} show the resulting posteriors for
the number of planets in the same two example datasets. On the left, $N_p=0$ has
the highest posterior probability, pointing to no planet being detected, while
in the second dataset, $N_p=1$ is the most probable value. Figure
\ref{fig:examples_PK} shows the joint posterior for the planetary minimum mass
and the orbital period, combining all values of $N_p$, together with the three
regions. The dataset shown on the right contained two simulated planets, within
$R_2$ and $R_3$. Only one of these signals is detected, showing as an
over-density on the posterior inside $R_2$.

\begin{figure}
	\centering
	\includegraphics[width=\hsize]{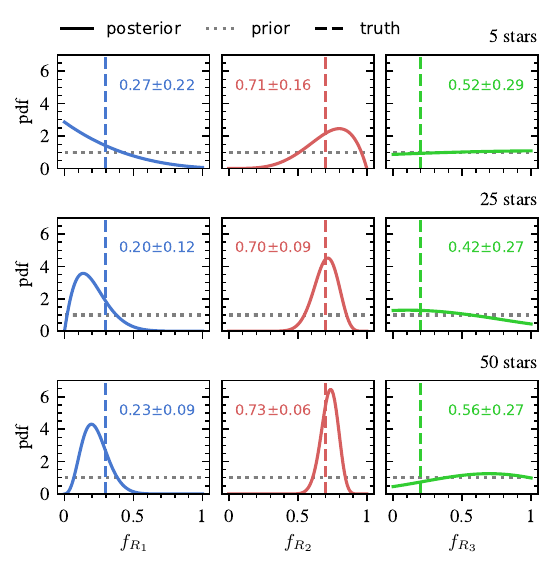}
	\caption{Occurrence rate estimates for the three regions $R_1$, $R_2$,
		$R_3$, from the analysis of 5, 25, and all 50 simulated datasets.
		The panels show the prior and posterior distributions (dotted and
		solid lines, respectively), and the true simulated values (vertical
		dashed lines). Each panel also shows the mean and standard
		deviation of the posterior distribution.}
	\label{fig:occurrence}
\end{figure}

In Fig.~\ref{fig:occurrence}, we show the occurrence rate estimates for the
three regions, as obtained by combining the analysis of 5, 25, and all 50
datasets. The plots show the assumed uniform prior and the resulting posterior
distribution for each \fR, as well as the true simulated values. It is clear
that the analysis of only 5 datasets provides little information about the
occurrence rates, even though it already assigns low posterior probability to
$\fR=0$ and $\fR=1$ in $R_2$, for example.

Including more stars improves the precision of the posterior estimates, but
differently for each region. This is discussed in more detail in
Appendix~\ref{sec:appC}. Since none of the simulated datasets is sensitive to
planets in $R_3$, with long orbital periods and small masses, the estimate for
the occurrence rate is uncertain, and the posterior is very close to the uniform
prior%
\footnote{The uniform prior has mean and standard deviation of 0.5 and 0.28.}%
\!.

\section{Discussion}
\label{sec:discussion}

We presented a method for estimating the occurrence rate of planets in a given
region of parameter space, from the analysis of RV data from several stars.
Using samples from the posterior distributions of the individual analyses, we
estimate the probability that at least one planet is present within a region
limited in orbital period and planetary mass. This only requires computing the
fraction of posterior samples that fall within the region, and comparing it to
the expected fraction from the prior.

Based on 50 simulated RV datasets, we inferred the occurrence rates for three
regions, defined in such a way as to probe different RV detection sensitivities.
In region $R_2$, all simulated signals should be easy to detect, even with only
$\sim$50 RV measurements (c.f. the example in Fig.~\ref{fig:examples}). On the
other hand, none of the simulated time series are sensitive to planet signals in
$R_3$. Region $R_1$ was chosen to represent an intermediate case, where some
signals may be detected.

We recovered unbiased estimates of the occurrence rates for the three regions.
As more stars are included, the posterior uncertainties decrease, and the
estimates become more precise (see Fig.~\ref{fig:occurrence}). For regions where
the RV data are insensitive to planet signals, the posterior remains close to
the prior, as expected.

This approach relies solely on existing posterior samples, does not require
injection-recovery tests, and bypasses the need to define an explicit detection
threshold. Moreover, the same set of posterior samples can be reused to estimate
occurrence rates for different regions or stellar subsamples, offering a
flexible and computationally efficient framework for population inference.

\subsection{Caveats and assumptions}

Listing the main assumptions of the proposed method may shed light on its
limitations. Violations of these assumptions may lead to underestimated
uncertainty or biased estimates of \fR.

	      To write Eqs. \eqref{eq:total_likelihood_0} and
	      \eqref{eq:total_likelihood}, we assumed that the RV datasets for each
	      star are independent (conditionally on knowing \fR). This may not be
	      true in real RV surveys, either because of shared instrumental
	      systematics or the presence of correlations within the stellar
	      population under study. For example, the instrument used to measure
	      the RVs may imprint a residual calibration error in all datasets, or
	      the true planet occurrence may depend (strongly) on stellar parameters
	      such as mass, metallicity, or age.

	      Our analyses assumed a simple white noise model and the simulated
	      datasets also only included white noise. Real RV data almost certainly
	      contain other sources of noise, such as instrumental systematics or
	      stellar activity signals. Fortunately, the method allows for more
	      complex noise models to be used in the individual analyses, which may
	      improve the sensitivity of the RV datasets to lower mass planets, and
	      mitigate the appearance of spurious signals.

	      We also assumed to have samples representative of the true posterior
	      distributions, even in regions of low probability. In practice, care
	      must be taken to ensure that the sampling algorithm is performing
	      correctly across the parameter space (see Appendix \ref{sec:appD} for
	      some diagnostics).

\subsection{Future work}

Our method provides a simple way to make inferences about the exoplanet
population. The natural next step is to apply it to real RV datasets from
ongoing or historical surveys. We also plan to investigate how instrumental
systematics and stellar activity influence the inferred occurrence rates, and to
mitigate these effects by introducing more flexible noise models in the
individual analyses. In addition, our approach can be extended to study planet
multiplicity and other orbital characteristics, either by adapting the
definition of the regions of interest or changing the statistical model (see
Appendix \ref{sec:appD} for details).

More generally, even though we focused on RV analysis in this work, similar
approaches could be applied to transit and direct imaging surveys \citep[see,
e.g.,][]{foreman-mackey2014, bowler2020}, enabling a unified view of the
demographics of exoplanets across multiple detection techniques.

\begin{acknowledgements}
	We thank the anonymous referee for constructive feedback that improved this
	manuscript. This research was funded by the Swiss National Science
	Foundation (SNSF), grant numbers 200020, 205010.
\end{acknowledgements}

\bibliographystyle{aa}
\bibliography{refs}

\begin{appendix}

	\section{Estimating \texorpdfstring{$\pi(\Ponly)$}{π(P)}}
	\label{sec:appA}

	The assumed priors for the model parameters define the prior distribution
	for the event \Ponly for each star: a Bernoulli distribution with
	probability $f_0$ (see Eq. \ref{eq:bernoulli_f0}). In the uninteresting
	limit where the priors exclude the region of interest, this probability is
	zero and the proposed method to estimate the occurrence rate is ill-defined.
	In other cases, the prior probability can be computed from the separable
	priors for the orbital period $\pi_P$, the semi-amplitude $\pi_K$, the
	eccentricity $\pi_e$, and the number of planets $\pi_{N_p}$.

	In practice, we estimate $f_0$ using Monte Carlo: we draw a large number of
	random samples from $\pi_P$, $\pi_K$, and $\pi_e$, calculate the planet mass
	for each sample using the stellar mass, and count the number of samples that
	fall within the region of interest. If the cumulative distribution function
	of $\pi_P$ is known, the random sampling can be made more efficient (and
	lower variance) by only sampling periods within the region. Denoting the
	fraction of samples falling within the region as $F$, we then calculate
	\begin{equation}\label{eq:mc_estimate}
		f_0 = 1 - \sum_{N_p=\,0}^{N_p^{\,\rm max}} \pi_{N_p} (N_p) \cdot ( 1 - F )^{N_p}
		,
	\end{equation}
	where $N_p^{\,\rm max}$ is the maximum value of $N_p$ (set to 5 in our
	analyses, see Table \ref{tab:priors}). This expression accounts for the ``at
	least one'' constraint on the definition of \Ponly. The calculation of $f_0$
	needs to be repeated for each star with the appropriate value for the
	stellar mass. With modern hardware, obtaining the random samples and
	calculating Eq. \eqref{eq:mc_estimate} can be achieved in about one second.

\begin{figure}[h]
		\centering
		\includegraphics[width=\hsize]{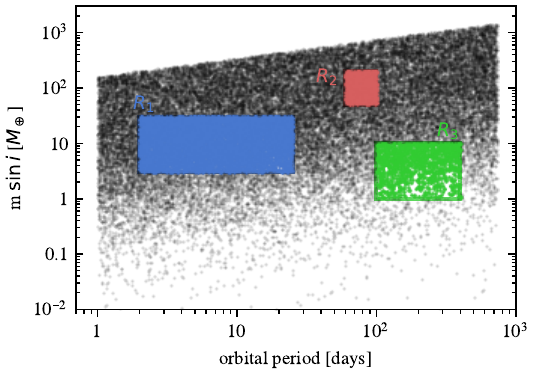}
		\caption{Procedure to estimate $f_0$ using random samples from the prior
			distribution that fall inside each region of interest.}
		\label{fig:f0}
	\end{figure}

	In Fig. \ref{fig:f0}, we sketch this calculation, showing several thousand
	random samples from the prior distribution (for $N_p=1$) in the plane of
	orbital period and planet minimum mass, together with the three regions
	$R_1$, $R_2$, and $R_3$ defined in the main text. For our choice of priors,
	the fraction of samples that fall within the regions and the corresponding
	values of $f_0$ are
	\begin{itemize}
		\item $F \approx 0.142$ and $f_0 \approx 0.295$ for $R_1$
		\item $F \approx 0.022$ and $f_0 \approx 0.053$ for $R_2$
		\item $F \approx 0.029$ and $f_0 \approx 0.072$ for $R_3$
	\end{itemize}
	as obtained from 10,000,000 prior samples.

	If the prior distribution is assigned directly to planet mass instead of RV
	semi-amplitude, there may be analytical expressions for all cumulative
	distributions, allowing for a calculation of $f_0$ that does not rely on
	Monte Carlo.

	\section{Model parameters and priors}
	\label{sec:appB}

	This appendix details the parameters and prior distributions used in the RV
	analysis. We parameterise each of the $N_p$ Keplerian functions with the
	orbital period $P$, the semi-amplitude $K$, the orbital eccentricity $e$,
	the argument of pericentre $\omega$, and the mean anomaly at the epoch
	$M_0$. The reference epoch is set to the time of the first observation. The
	systemic velocity is denoted as $v_{\rm sys}$ and is fixed to zero for
	simplicity. A jitter parameter $\sigma_{\!J}$ is used to account for
	otherwise unmodelled white noise.

\begin{figure}[h]
		\centering
		\includegraphics[width=0.9\hsize]{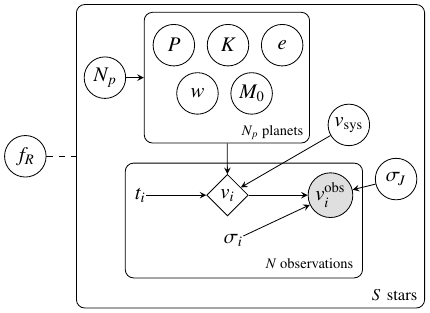}
		\caption{
			Probabilistic graphical model for the RV analysis.
			An arrow between two nodes indicates the direction of conditional
			dependence. The circled nodes are parameters of the model, whose
			joint distribution is sampled. The filled node represents the
			observed RVs. The $t_i$ and $\sigma_i$ nodes are assumed given and
			thus fixed, while the $v_i$ are deterministic functions of other
			nodes. Variables inside boxes are repeated a given number of times.
			The dashed line connecting \fR indicates the importance sampling
			scheme used to estimate the posterior for this parameter. }
		\label{fig:pgm}
	\end{figure}

	Each dataset consists of $N$ observations containing the times $t_i$, the
	radial-velocity measurements $v^{\rm obs}_i$, and the associated
	uncertainties $\sigma_i$. We further define the model radial velocities,
	$v_i$, as deterministic functions of the orbital parameters and $v_{\rm
	sys}$. The connections between all the parameters are shown in Fig.
	\ref{fig:pgm} as a probabilistic graphical model.

	\noindent
	Prior distributions for each parameter are listed in Table~\ref{tab:priors}.
	Note that these same priors were used for sampling the orbital parameters of
	the simulated planets, while ensuring the values fell within each region of
	interest.

	\begin{table}[h]
		\renewcommand{\arraystretch}{1.15}
		\caption{Prior distributions used in the RV analysis.}
		\label{tab:priors}
		\centering
		\begin{tabular}{l c l l}
			\hline\hline
			Parameter           & Units & Prior                                                    \\
			\hline
			$v_{\rm sys}$       & \ms   & $\mathcal{F}$\,(0)                                       \\
			$\sigma_j$ (jitter) & \ms   & $\mathcal{H}\mathcal{G}$\,(10)                           \\
			%
			$N_p$               &       & $\mathcal{U}_d \left( 0, 5 \right)$                      \\
			$P$                 & days  & $\mathcal{LU}  \left( 1\,\text{day}, 2\,\text{yr} \right)$ \\
			$K$                 & \ms   & $\mathcal{MLU} \left( 1, 100 \right)$                    \\
			$e$                 &       & $\mathcal{K}   \left( 0.867, 3.03 \right)$               \\
			$\omega$            & rad   & $\mathcal{U}   \left( 0, 2\pi \right)$                   \\
			$M_0$               & rad   & $\mathcal{U}   \left( 0, 2\pi \right)$                   \\
			\hline
		\end{tabular}
		\tablefoot{Distributions are abbreviated as follows:
			$\mathcal{F}$~=~fixed;
			$\mathcal{H}\mathcal{G}$~=~Half-Gaussian with standard deviation;
			$\mathcal{U}$~=~uniform with lower and upper limits; subscript $d$
			indicates a discrete distribution;
			$\mathcal{LU}$~=~log-uniform with lower and upper limits;
			$\mathcal{MLU}$~=~modified log-uniform with knee and upper limit;
			$\mathcal{K}$~=~Kumaraswamy with parameters $a$ and $b$.
		}
		\renewcommand{\arraystretch}{1}
	\end{table}

	\section{Posterior uncertainties}
	\label{sec:appC}

	The usual method to estimate planet occurrence rates from RV surveys relies
	on binomial statistics. That is, it is assumed that the number of planets
	detected, $k$, follows a binomial distribution with parameters $S$ and \fR,
	using our notation. To account for the different sensitivity of each star,
	the average completeness within the region $R$ is used to calculate an
	``effective'' number of stars, $S_{\!\rm eff}$, that replaces $S$ in the
	binomial probability.

	Even if not explicitly stated, the inference for \fR within this context is
	usually based on the beta-binomial distribution: a beta distribution (of
	which the uniform distribution is a special case) is often assumed as the
	prior for \fR. The beta is a conjugate prior to the binomial, so the
	posterior distribution for \fR is also a beta distribution, with parameters
	\begin{equation*}
		\alpha = k + 1 \quad \text{and} \quad \beta = S - k + 1
	\end{equation*}
	where the $+1$ factors come from the uniform prior.

\begin{figure}[h]
		\centering
		\includegraphics[width=\hsize]{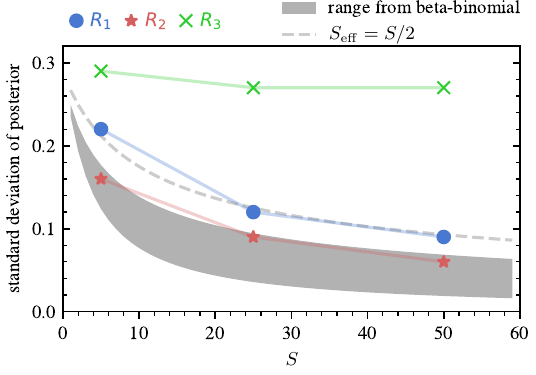}
		\caption{Standard deviation of the posterior distribution for \fR as a
			function of the number of stars in the sample $S$.}
		\label{fig:beta_std}
	\end{figure}

	The standard deviation of the posterior for \fR is maximized if there are
	$\lfloor S/2 \rfloor$ or $\lceil S/2 \rceil$ detections, it is minimized if
	there are zero or $S$ detections, and depends on the true value of \fR
	within this range. This is shown in Fig. \ref{fig:beta_std}, as a function
	of $S$. As we increase the number of stars in the sample, the posterior
	uncertainty for \fR will stay within this range. If only $S_{\!\rm eff}$
	stars have enough sensitivity to detect a planet, the uncertainty in the
	posterior is larger. Figure \ref{fig:beta_std} shows a dashed curve for the
	case where $S_{\!\rm eff} = S / 2$ (completeness of 50\%).

	Overplotting the standard deviations of the inferred posterior distributions
	for \fR using our method (see values listed in Fig.~\ref{fig:occurrence})
	shows good agreement with the beta-binomial results. We see two sources of
	uncertainty being reflected in the posterior: from the number of stars
	included in the analysis and from the sensitivity of each star's RV
	timeseries.

	As expected, each of the three regions we studied is affected differently by
	these uncertainties. All RV datasets are sensitive to planets in $R_2$, so
	the posterior uncertainty comes only from the limited number of stars in the
	sample. For $R_1$, only a fraction of the stars are sensitive to planets
	with these orbital periods and masses, so the posterior uncertainty is
	larger, and comparable to a completeness fraction of 50\%. Inference in
	region $R_3$ is mostly controlled by the prior.

	In summary, the uncertainty in our inferred occurrence rates is compatible
	with other methods, but our approach completely avoids having to determine
	$k$ and the survey completeness.

	\section{Generalizations and diagnostics}
	\label{sec:appD}

	In section \ref{subsec:occurrence}, the planet occurrence rate was defined
	as the fraction of stars with at least one planet within a region of
	parameter space. An alternative definition could be the average number of
	planets per star within that region \citep[e.g.][]{zhu2019}. With some
	modifications, our method can be used to estimate this quantity.

	We define the event ``there are $n_p$ planets with parameters in a given
	region'', denoted by $n_p$ itself (not to be confused with $N_p$), which
	takes non-negative integer values. For each star, we assume this event
	follows a Poisson distribution with rate parameter equal to the average
	number of planets per star within the region, here denoted as $\lambda_R$:
	\begin{equation}\label{eq:poisson}
      p(\lambda_R; n_p) = \frac{ \lambda_R^{n_p} \, e^{-\lambda_R} }{ n_p! }
  	\end{equation}
	The priors used in the individual analyses again define a prior probability
	for $n_p$, but now it is not within the same distribution family. The
	original priors were defined for the total number of planets in the system,
	$N_p$. The prior probability that $n_p$ out of $N_p$ planets are within $R$
	is given by a binomial distribution. After marginalising out $N_p$, we
	obtain:
	\begin{equation}
		\pi(F; n_p) = \sum_{N_p=0}^{5} \pi_{N_p}(N_p) \cdot 
					\binom{N_p}{n_p} \cdot F^{n_p} \cdot (1 - F)^{N_p - n_p}
	\end{equation}
	where $F$ is the probability of one planet being within $R$, which was
	estimated using Monte Carlo (see Appendix \ref{sec:appA}). The upper limit
	in the sum is set by the maximum value for $N_p$ in the prior.

	The ratio $p(\lambda_R; n_p) / \pi(F; n_p)$ can then be used to compute the
	likelihood, analogously to Eq. \eqref{eq:is_likelihood} but using $n_{p,jk}$
	as the number of planets in $R$ for each posterior sample, and thus the
	posterior for $\lambda_R$ can be evaluated after defining a prior
	$p(\lambda_R)$. As before, a uniform distribution, from 0 to 5, may be a
	suitable choice.

	Some caveats are worth mentioning. The Poisson assumption \eqref{eq:poisson}
	implies that different planets around the same star occur independently,
	which is likely not true. Moreover, the original prior $\pi(F; n_p)$ and the
	new prior $p(\lambda_R; n_p)$ may be very different for some values of
	$\lambda_R$, leading to high variance in the importance sampling
	approximation (Eq. \ref{eq:is}). 
	
	\begin{figure*}
		\centering
		\includegraphics[width=0.9\hsize]{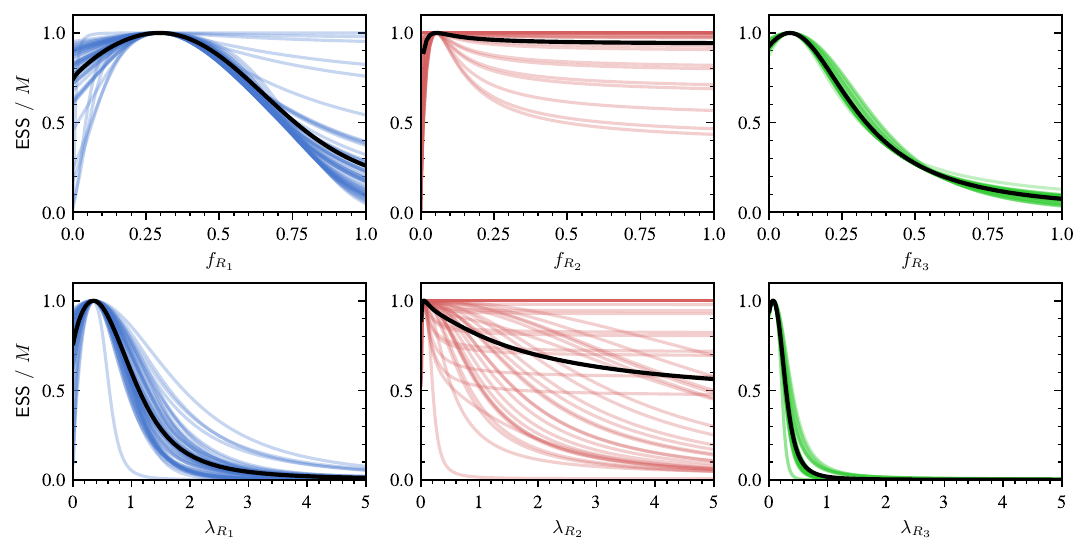}
		\caption{Ratio of effective sample size (ESS) in the importance sampling
				approximation to total number of posterior samples versus
				fraction of stars with planets \fR (top) and average number of
				planets per star $\lambda_R$ (bottom), for each star and each
				box. The thick black curves show the mean ratio across all
				stars.}
		\label{fig:ess}
	\end{figure*}

	\begin{figure*}
		\centering
		\includegraphics[width=0.9\hsize]{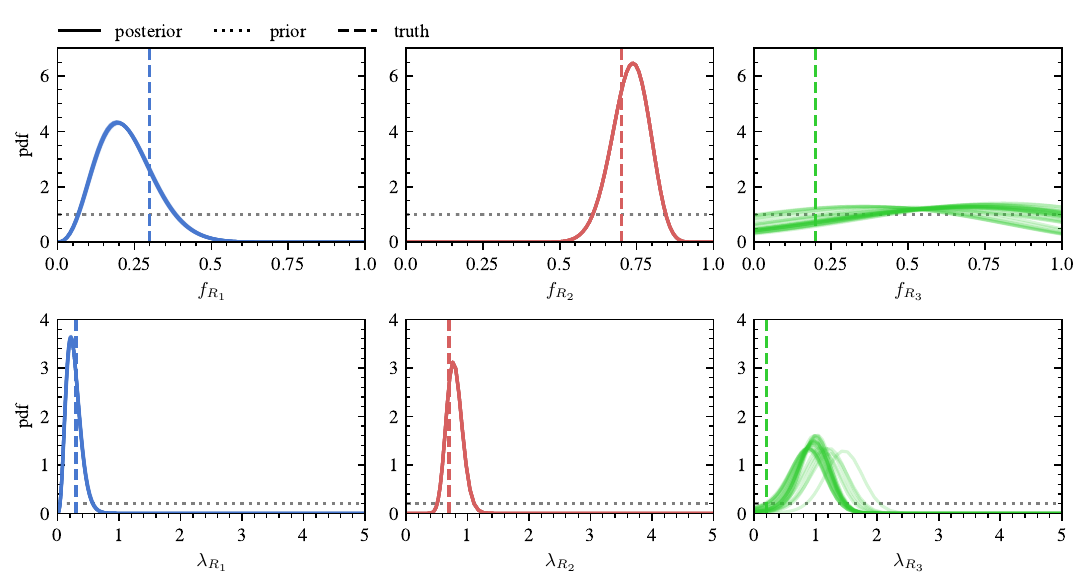}
		\caption{Results from bootstrap resampling of the posterior samples. The
                 panels show the posterior distributions for the fraction of
                 stars with planets (top) and the average number of planets per
                 star (bottom) for the three regions $R_1$, $R_2$, $R_3$, and
                 from the analysis of all 50 simulated datasets. The lines are
                 the same as in Fig.~\ref{fig:occurrence}. Note that the true
                 values of \fR and $\lambda_R$ are the same due to the way the
                 simulated datasets were generated.}
		\label{fig:bootstrap}
	\end{figure*}

	A useful diagnostic when using importance sampling is the effective sample
	size (ESS), which can be computed from the ratio of the two priors, usually
	called the importance weight \citep[e.g.][]{elvira2022}. The ESS essentially
	quantifies how many of the posterior samples, obtained based on
	$\pi(f_0;\Ponly)$ or $\pi(F; n_p)$, have substantial weight under the new
	priors $p(\fR;\Ponly)$ or $p(\lambda_R; n_p)$. We estimate the ESS for each
	simulated star as
	\begin{equation}
		\text{ESS}_j = \frac{ \left( \sum_{k=1}^{M_j} w_{jk} \right)^2 }{ \sum_{k=1}^{M_j} w_{jk}^2 }
	\end{equation}
	with the weights defined as
	\begin{equation}
		w_{jk} = \frac{ p(\fR; \Ponly_{jk}) }{ \pi(\Ponly_{jk}) }
		\quad \text{or} \quad
		w_{jk} = \frac{ p(\lambda_R; n_{p,jk}) }{ \pi(n_{p,jk}) }
		.
	\end{equation}

	Figure \ref{fig:ess} shows the ratio of the ESS to the total number of
	posterior samples for each star and region, as a function of \fR and
	$\lambda_R$. In the case of \fR, the ESS is above 50\% of the number of
	posterior samples, except for values close to 1 for $R_1$ and above 0.25 for
	$R_3$. Therefore, the importance sampling approximation is less accurate in
	these cases. For $R_1$, this has a small impact on the results since the
	posterior for \fR peaks at lower values. The ESS is in general smaller for
	$\lambda_R$, especially for $R_3$, due to the larger mismatch between the
	original and new priors.
	
	Another way to diagnose the results is to perform bootstrap resampling, by
	randomly shuffling (with replacement) the posterior samples for each star
	and recomputing the occurrence rate posteriors. An example of this is shown
	in Fig. \ref{fig:bootstrap}, which plots the posteriors for \fR and
	$\lambda_R$ for each region, obtained from 30 resampling iterations. The
	posteriors show little variation for $R_1$ and $R_2$, but the variance and
	bias are higher for $R_3$.

	Finally, we note that a perhaps more natural generalization of the method
	would be to estimate the individual probabilities for $n_p=0$, $n_p=1$, and
	so on, while maybe collapsing all values above a certain number into one bin
	(e.g., $n_p \geq 2$). These individual probabilities could be modelled with
	a Dirichlet distribution, allowing for a larger flexibility than the
	Poisson, with few additional parameters. Such a parameterisation might also
	be more easily compared to empirical probabilities provided by planet
	formation models.

\end{appendix}

\end{document}